\documentclass[11pt,twoside]{article}

\usepackage{asp2014}

\aspSuppressVolSlug
\resetcounters

\begin{document}

\title{Deep Learning for AGILE Anticoincidence System's Background Prediction from Orbital and Attitude Parameters}

\author{N.~Parmiggiani,$^1$ A.~Bulgarelli,$^1$ A.~Macaluso,$^2$ A.~Ursi,$^{3,4}$, L.~Castaldini,$^1$ A.~Di~Piano,$^{5,1}$ R.~Falco,$^{6,1}$ V.~Fioretti,$^1$ G.~Panebianco,$^{7,1}$ C.~Pittori,$^{8,9}$ and M.~Tavani$^4$ }

\affil{$^1$INAF OAS Bologna, Via P. Gobetti 93/3, 40129 Bologna, Italy. \email{nicolo.parmiggiani@inaf.it}}

\affil{$^2$German Research Center for Artificial Intelligence (DFKI), 66123 Saarbruecken, Germany.}

\affil{$^3$ASI, via del Politecnico snc, I-00133 Roma (RM), Italy.
}

\affil{$^4$INAF/IAPS Roma, Via del Fosso del Cavaliere 100, I-00133 Roma, Italy.}

\affil{$^5$Universit\`{a} degli studi di Modena e Reggio Emilia, DIEF, Via Pietro Vivarelli 10, 41125 Modena, Italy
}

\affil{$^6$Computer Science and Engineering - DISI, University of Bologna, Via Zamboni 33, Bologna, Italy
}

\affil{$^7$Department of Physics and Astronomy, University of Bologna, Via Gobetti 93/2, 40129, Bologna, Italy
}

\affil{$^8$INAF/OAR Roma, Via Frascati 33, I-00078 Monte Porzio Catone, Roma, Italy.}

\affil{$^9$ASI/SSDC Roma, Via del Politecnico snc, I-00133 Roma, Italy.}

\paperauthor{Nicol\`{o}~Parmiggiani}{nicolo.parmiggiani@inaf.it}{0000-0002-4535-5329}{INAF}{OAS}{Bologna}{BO}{40129}{Italy}
\paperauthor{Andrea~Bulgarelli}{andrea.bulgarelli@inaf.it}{0000-0001-6347-0649}{INAF}{OAS}{Bologna}{BO}{40129}{Italy}
\paperauthor{Alessandro~Ursi}{alessandro.ursi@inaf.it}{0000-0002-7253-9721}{INAF}{IAPS}{Roma}{RO}{00133}{Italy}
\paperauthor{Antonio~Macaluso}{antonio.macaluso.90@gmail.com }{0000-0002-1348-250X}{}{DFKI}{Saarbruecken}{66123}{}{Germany}
\paperauthor{Luca~Castaldini}{luca.castaldini@inaf.it}{0009-0000-5501-4328}{INAF}{OAS}{Bologna}{BO}{40129}{Italy}
\paperauthor{Riccardo~Falco}{riccardo.falco@inaf.it}{0009-0004-1676-7596}{INAF}{OAS}{Bologna}{BO}{40129}{Italy}
\paperauthor{Ambra~Di~Piano}{ambra.dipiano@inaf.it}{0000-0002-9894-7491}{INAF}{OAS}{Bologna}{BO}{40129}{Italy}
\paperauthor{Valentina~Fioretti}{valentina.fioretti@inaf.it}{0000-0002-6082-5384}{INAF}{OAS}{Bologna}{BO}{40129}{Italy}
\paperauthor{Gabriele~Panebianco}{gabriele.panebianco@inaf.it}{0000-0002-3410-8613}{INAF}{OAS}{Bologna}{BO}{40129}{Italy}
\paperauthor{Carlotta~Pittori}{carlotta.pittori@inaf.it}{0000-0001-6661-9779}{INAF}{OAR}{Monte Porzio Catone}{RO}{00078}{Italy}
\paperauthor{Marco~Tavani}{marco.tavani@inaf.it}{0000-0003-2893-1459}{INAF}{OAR}{Monte Porzio Catone}{RO}{00078}{Italy}



\begin{abstract}
AGILE is an Italian Space Agency (ASI) space mission launched in 2007 to study X-ray and gamma-ray phenomena in the energy range from $\sim$20 keV to $\sim$10 GeV. The AGILE AntiCoincidence System (ACS) detects hard-X photons in the 50 - 200 keV energy range and continuously stores each panel's count rates in the telemetry. We developed a new Deep Learning (DL) model to predict the background of the AGILE ACS top panel using the satellite's orbital and attitude parameters. This model aims to learn how the orbital and spinning modulations of the satellite impact the background level of the ACS top panel. The DL model executes a regression problem, and is trained with a supervised learning technique on a dataset larger than twenty million orbital parameters' configurations. Using a test dataset, we evaluated the trained model by comparison of the predicted count rates with the real ones. The results show that the model can reconstruct the background count rates of the ACS top panel with an accuracy of 96.7\%, considering the orbital modulation and spinning of the satellite. Starting from these promising results, we are developing an anomaly detection method to detect Gamma-ray Bursts when the differences between predicted and real count rates exceed a predefined threshold.

\end{abstract}



\section{Introduction}

AGILE (Astrorivelatore Gamma ad Immagini LEggero - Light Imager for Gamma-Ray Astrophysics) is a space mission of the Italian Space Agency (ASI) devoted to high-energy astrophysics \citep{2008NIMPA.588...52T, BULGARELLI2010213}, launched in 2007 and still operational. Since 2009, the AGILE satellite continuously spins around its sun-pointing axis, with an angular speed of about 0.8 degrees/sec, thus completing a rotation every ~7 minutes. This work uses data acquired during the so-called "spinning mode" observing period. AGILE has an anti-coincidence system (ACS, \citet{2006NIMPA.556..228P}) comprising five independent panels surrounding all AGILE detectors, used to efficiently reject background-charged particles. The ACS detects hard-X photons in the 50 - 200 keV energy range and continuously stores each panel's count rates in the telemetry as ratemeters data, with 1.024 s resolution. We developed a new Deep Learning (DL, \citet{Goodfellow-et-al-2016}) model to predict the background value of the AGILE ACS top panel, using the satellite's orbital parameters. We decided to analyze the data of this panel because it is perpendicular to the pointing direction of the payload detectors. A similar method is presented in \citet{2023ExA...tmp...59C}. We aim to use the trained DL model as a reliable method for detecting Gamma-ray bursts (GRB). Indeed we can detect GRBs when the real count rates is over a certain threshold with respect to the prediction. This new detection method can be used to analyze the AGILE data archive or be implemented inside the AGILE real-time analysis system \citep{Bulgarelli_2014,2023A&C....4400726P} for the follow-up of transient events.

\section{Deep Learning Model}

The DL model executes a regression problem and is implemented with a dense neural network of three hidden layers \citep{Goodfellow-et-al-2016}. The first layer has 1024 neurons, while the last two layers have 512 neurons. The three hidden layers use the Relu activation function, while the output layer uses a linear activation function to provide a continue value as output. We used dropout layers with a dropout rate of 3\% as a regularization mechanism. We implemented the model using two open-source frameworks: Keras \footnote{https://keras.io/} and Tensorflow \footnote{https://www.tensorflow.org/}.

The input of the model is the AGILE orbital parameters' configuration, and the output is the predicted count rate of the ACS top panel. The model is trained with a supervised technique thus, we created a labeled dataset to train and test the model.

The dataset contains more than twenty million orbital parameters' configurations extracted from the 2020 data archive with the associated ACS top panel count rates (the labels). The orbital configuration parameters are Earth Latitude and Longitude, RA-Dec of the ACS top panel attitude, Earth RA-Dec, seconds from the start of the day and day of the year. We split the dataset into training and test datasets, with respective percentages of 90\% and 10\%. The orbital parameters and the ACS top ratemeters are then normalized between 0 and 1 to improve the training process. 

From the dataset we excluded time windows containing the South Atlantic Anomalies passages and known GRBs, as to analyze background-only data. The list of known GRBs is taken from the public GRBweb catalog\footnote{https://heasarc.gsfc.nasa.gov/FTP/fermi/data/gbm/bursts/} \citep{PhysRevD.102.103014} that collects GRBs detected from several facilities. In addition, we considered the GRBs detected by AGILE. The training uses the Nadam optimizer and the Mean Absolute Error loss function. The learning rate of the optimizer is variable as follows: i) 0.01 for epochs lower than five, ii) 0.006 for epochs between five and ten, and iii) 0.0001 for epochs greater than ten. We trained the model for 109 epochs with a batch size of over 130 thousand orbital parameters using an NVIDIA Tesla V100 GPU on a Power 9 processor.

\section{Results}
 
We evaluated the trained model using the test dataset, which contains more than two hundreds thousands orbital parameters' configurations and we compared the predicted ACS top panel count rates with the real ones. Fig. \ref{fig1} shows an example of two ACS top time series with real (blue) and predicted count rates (red). The model learned the orbital and spinning modulations and can accurately predict the background count rates of the ACS top panel in each orbital and spinning phase. The mean reconstruction accuracy is 96.7\%. This result is impressive considering the complexity of the satellite's behavior and the variability of the data. 

\articlefiguretwo{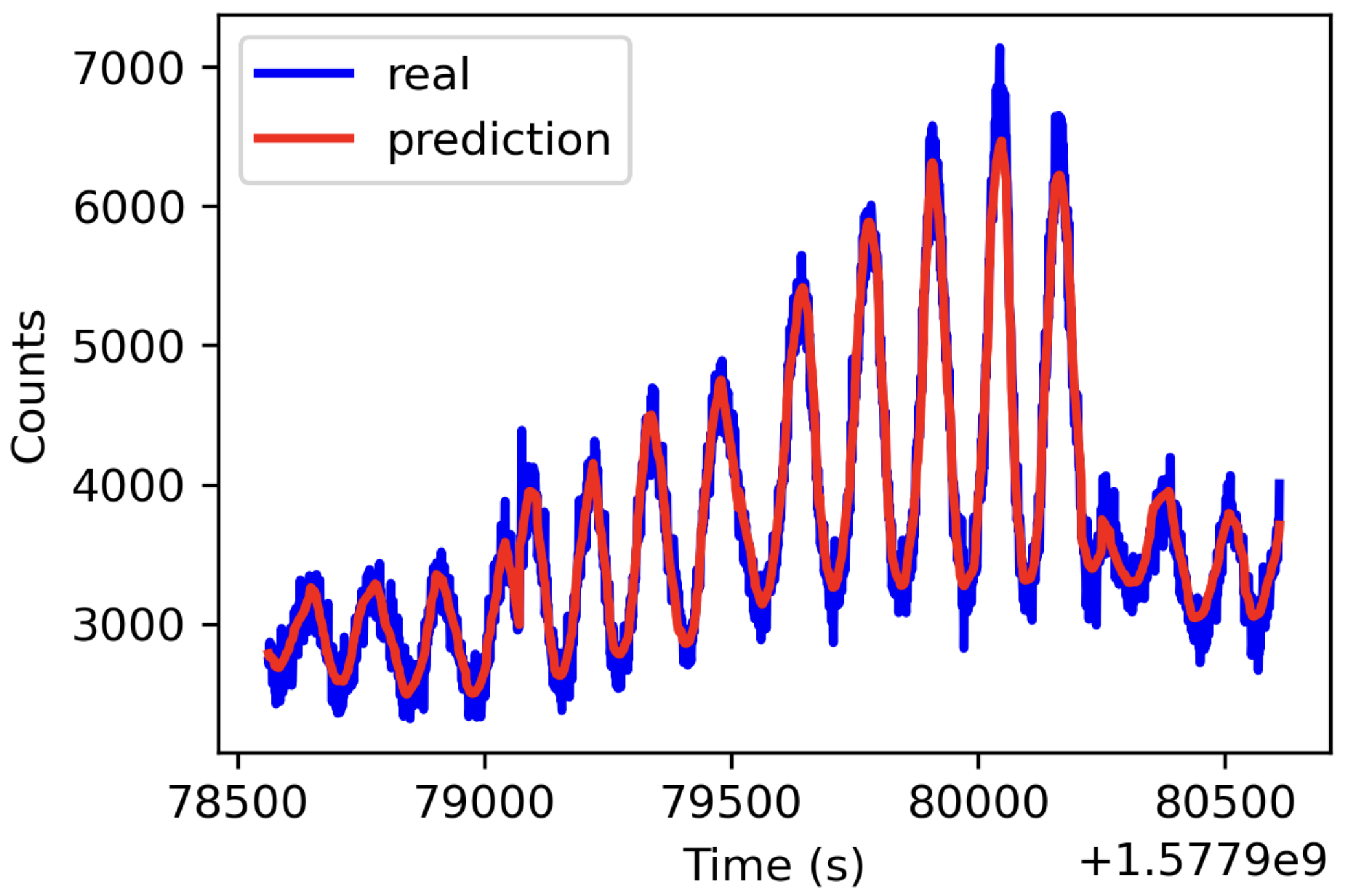}{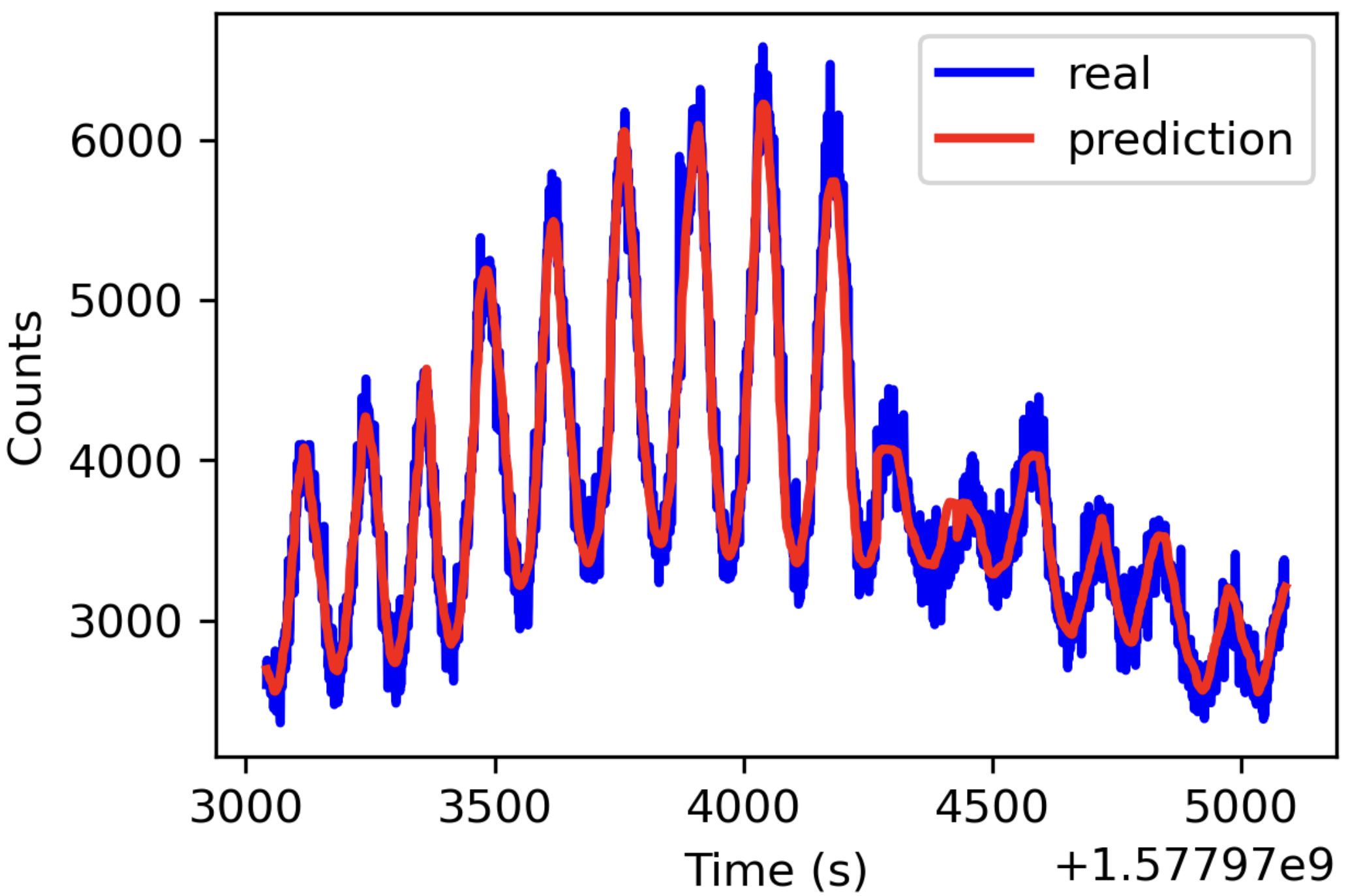}{fig1}{Two examples of ACS top light curves. The blue line represents the real count rates while the red line represents the count rates predicted by the DL model.}

\section{Conclusion and future works}

We developed a deep neural network trained with millions of AGILE orbital parameters' configurations to perform a regression task predicting the background count rates of the AGILE Anticoincidence System. The results show that this model can predict the ACS top count rates with an accuracy of 96.7\%. Starting from these promising results, we are developing an anomaly detection method to detect Gamma-Ray Bursts when the differences between predicted and real count rates in the ACS top ratemeters (Fig. \ref{fig2}) exceed a predefined threshold. We plan to use the p-value distribution of the count rate differences to calculate the statistical significance of the detected GRBs. Compared to existing methods such as \citet{2023ApJ...945..106P}, the main advantage of this method is that it can analyze the ACS top data without using a detrending algorithm, which can introduce artificial anomalies in the data. The presented method can be used to train DL models to analyze the data from other AGILE detectors or the data acquired by new next-generation high-energy space missions such as COSI \citep{Tomsick:2023Xz}.

\articlefiguretwo{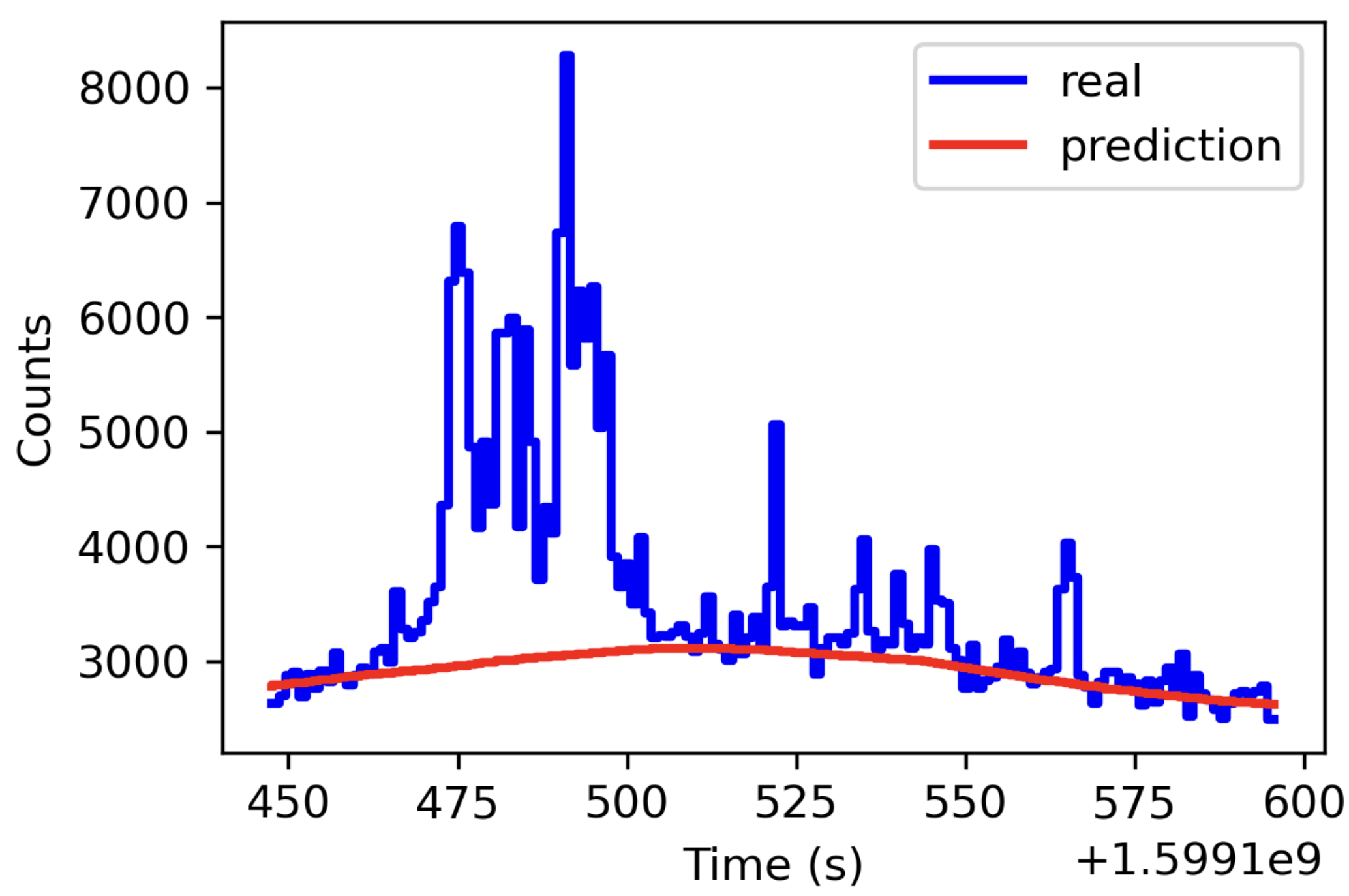}{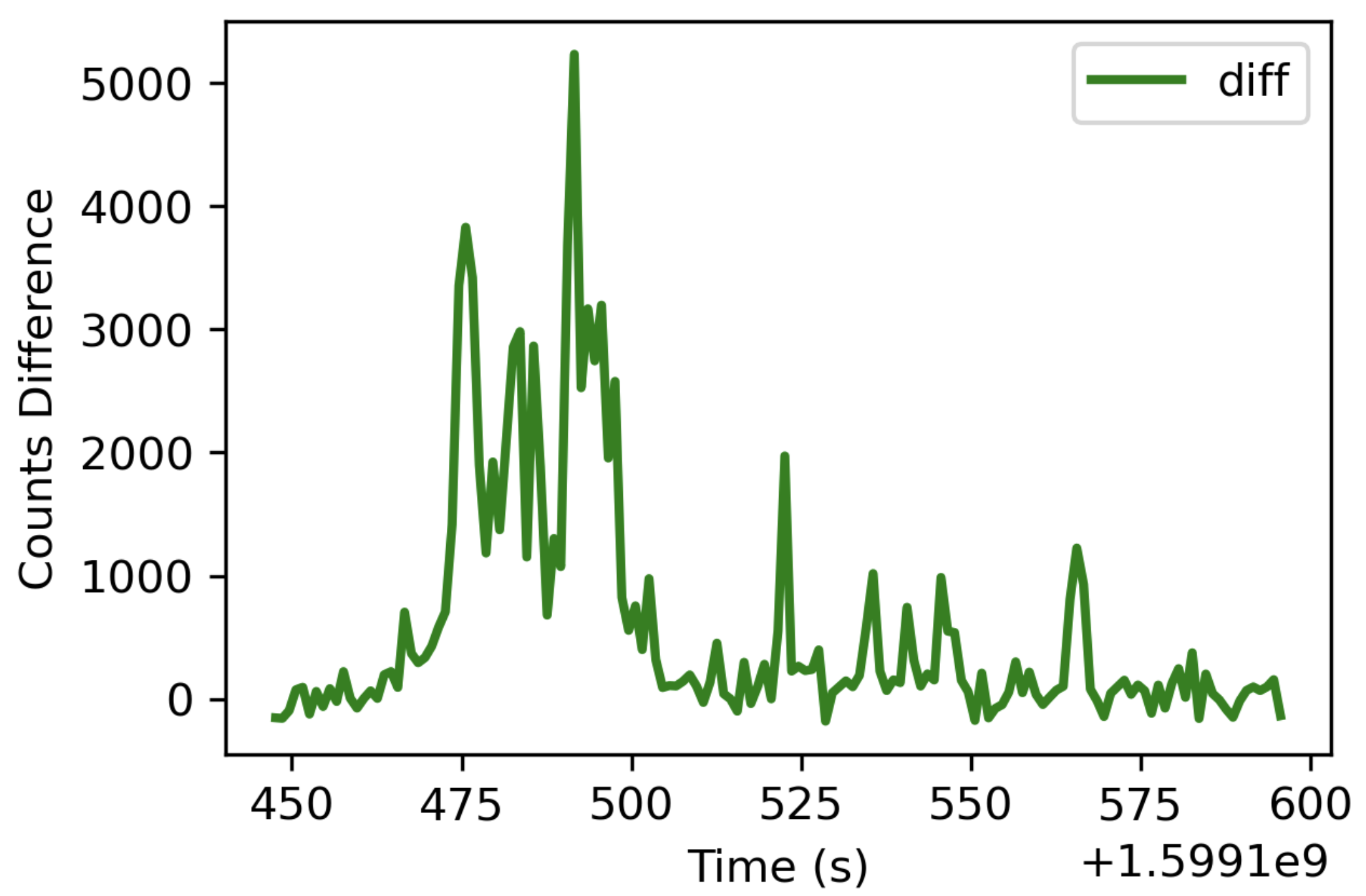}{fig2}{Light curve of the ACS top count rates during GRB200903E.  \emph{Left:} The red line represent the count rates predicted using the orbital parameters of AGILE, while the blue line represents the real count rates detected by the instruments. \emph{Right:} The green line represents the differences between the predicted and real count rates.}

\acknowledgements The AGILE Mission is funded by the Italian Space Agency (ASI) with scientific and programmatic participation by the Italian National Institute for Astrophysics (INAF) and the Italian National Institute for Nuclear Physics (INFN). The investigation is supported by the ASI grant I/028/12.7-2022 and also partially funded with an INAF "Mini-Grant" 2022. We thank the ASI management for unfailing support during AGILE operations. We acknowledge the effort of ASI and industry personnel in operating the ASI ground station in Malindi (Kenya), the Telespazio Mission Control Center at Fucino, and the data processing done at the ASI/SSDC in Rome: the success of AGILE scientific operations depends on the effectiveness of the data flow from Kenya to SSDC and the data analysis and software management. We thank IBM Italy and Var Group S.p.A. for the opportunity to use a Power 9 with NVIDIA Tesla V100.

\bibliography{P410}  

\begin{thebibliography}{}
\expandafter\ifx\csname natexlab\endcsname\relax\def\natexlab#1{#1}\fi
\expandafter\ifx\csname url\endcsname\relax
  \def\url#1{\texttt{#1}}\fi
\expandafter\ifx\csname urlprefix\endcsname\relax\def\urlprefix{URL }\fi
\providecommand{\eprint}[2][]{\url{#2}}

\bibitem[{{Bulgarelli} et~al.(2010)}]{BULGARELLI2010213}
{Bulgarelli}, A., et~al. 2010, Nuclear Instruments and Methods in Physics Research Section A: Accelerators, Spectrometers, Detectors and Associated Equipment, 614, 213. \urlprefix\url{https://www.sciencedirect.com/science/article/pii/S0168900209023882}

\bibitem[{{Bulgarelli} et~al.(2013)}]{Bulgarelli_2014}
--- 2013, The Astrophysical Journal, 781, 19. \urlprefix\url{https://dx.doi.org/10.1088/0004-637X/781/1/19}

\bibitem[{Coppin et~al.(2020)Coppin, de~Vries, \& van Eijndhoven}]{PhysRevD.102.103014}
Coppin, P., de~Vries, K.~D., \& van Eijndhoven, N. 2020, Phys. Rev. D, 102, 103014. \urlprefix\url{https://link.aps.org/doi/10.1103/PhysRevD.102.103014}

\bibitem[{{Crupi} et~al.(2023){Crupi}, {Dilillo}, {Bissaldi}, {Ward}, {Fiore}, \& {Vacchi}}]{2023ExA...tmp...59C}
{Crupi}, R., {Dilillo}, G., {Bissaldi}, E., {Ward}, K., {Fiore}, F., \& {Vacchi}, A. 2023, Experimental Astronomy. \eprint{2303.15936}

\bibitem[{Goodfellow et~al.(2016)Goodfellow, Bengio, \& Courville}]{Goodfellow-et-al-2016}
Goodfellow, I., Bengio, Y., \& Courville, A. 2016, Deep Learning (MIT Press). \url{http://www.deeplearningbook.org}

\bibitem[{{Parmiggiani} et~al.(2023{\natexlab{a}})}]{2023ApJ...945..106P}
{Parmiggiani}, N., et~al. 2023{\natexlab{a}}, \apj, 945, 106

\bibitem[{{Parmiggiani} et~al.(2023{\natexlab{b}})}]{2023A&C....4400726P}
--- 2023{\natexlab{b}}, Astronomy and Computing, 44, 100726

\bibitem[{{Perotti} et~al.(2006){Perotti}, {Fiorini}, {Incorvaia}, {Mattaini}, \& {Sant'Ambrogio}}]{2006NIMPA.556..228P}
{Perotti}, F., {Fiorini}, M., {Incorvaia}, S., {Mattaini}, E., \& {Sant'Ambrogio}, E. 2006, Nuclear Instruments and Methods in Physics Research A, 556, 228

\bibitem[{{Tavani} et~al.(2008)}]{2008NIMPA.588...52T}
{Tavani}, M., et~al. 2008, Nuclear Instruments and Methods in Physics Research A, 588, 52

\bibitem[{{Tomsick} et~al.(2023)}]{Tomsick:2023Xz}
{Tomsick}, J., et~al. 2023, PoS, ICRC2023, 745

\end{thebibliography}


\end{document}